\documentclass[conference, anonymous]{IEEEtran}
\IEEEoverridecommandlockouts
\usepackage{cite}
\usepackage{amsmath,amssymb,amsfonts}
\usepackage{algorithmic}
\usepackage{graphicx}
\usepackage{textcomp}
\usepackage{xcolor}
\usepackage{bbding}
\usepackage{multirow}
\usepackage{tabularx}
\usepackage{threeparttable}
\usepackage{pgfplots}
\pgfplotsset{compat=1.18} 
\usetikzlibrary{patterns}
\usepackage{adjustbox} 
\usepackage{tikz}
\def\BibTeX{{\rm B\kern-.05em{\sc i\kern-.025em b}\kern-.08em
    T\kern-.1667em\lower.7ex\hbox{E}\kern-.125emX}}
\DeclareRobustCommand*{\ieeeauthorrefmark}[1]{%
    \raisebox{0pt}[0pt][0pt]{\textsuperscript{\footnotesize\ensuremath{#1}}}}

\begin{document}

\title{ASD-Chat: An Innovative Dialogue Intervention System for Children with Autism based on LLM and VB-MAPP\\

}

\author{
    \IEEEauthorblockN{
        Chengyun Deng\ieeeauthorrefmark{1,4\dag},
        Shuzhong Lai\ieeeauthorrefmark{1,5\dag},
        Chi Zhou\ieeeauthorrefmark{1,4},
        Mengyi Bao\ieeeauthorrefmark{6},\\
        Jingwen Yan\ieeeauthorrefmark{7},
        Haifeng Li\ieeeauthorrefmark{6},
        Lin Yao\ieeeauthorrefmark{1,2,4}\ieeeauthorrefmark{*},
        and Yueming Wang\ieeeauthorrefmark{3,4}
    }
    \IEEEauthorblockA{\ieeeauthorrefmark{1}MOE Frontiers Science Center for Brain and Brain-Machine Integration, Zhejiang University, Hangzhou, China}
    \IEEEauthorblockA{\ieeeauthorrefmark{2}Department of Neurobiology, Affiliated Mental Health Center \& Hangzhou Seventh People’s Hospital, \\
    Zhejiang University School of Medicine, Zhejiang University, Hangzhou, China}
    \IEEEauthorblockA{\ieeeauthorrefmark{3}Qiushi Academy for Advanced Studies (QAAS), Zhejiang University, Hangzhou, China}
    \IEEEauthorblockA{\ieeeauthorrefmark{4}College of Computer Science, Zhejiang University, Hangzhou, China}
    \IEEEauthorblockA{\ieeeauthorrefmark{5}Polytechnic Institute, Zhejiang University, Hangzhou, China}
    \IEEEauthorblockA{\ieeeauthorrefmark{6}Children's Hospital of Zhejiang University School of Medicine, Hangzhou, China}
    \IEEEauthorblockA{\ieeeauthorrefmark{7}Beijing IngCare Technology Co.Ltd., Beijing, China}
}

\maketitle
\footnote{\dag: Equal Contribution.}
\footnote{*: Corresponding Author. Email: lin.yao@zju.edu.cn}
\begin{abstract}
Early diagnosis and professional intervention can help children with autism spectrum disorder (ASD) return to normal life. However, the scarcity and imbalance of professional medical resources currently prevent many autistic children from receiving the necessary diagnosis and intervention. Therefore, numerous paradigms have been proposed that use computer technology to assist or independently conduct ASD interventions, with the aim of alleviating the aforementioned problem. However, these paradigms often lack a foundation in clinical intervention methods and suffer from a lack of personalization. Addressing these concerns, we propose ASD-Chat, a social intervention system based on VB-MAPP (Verbal Behavior Milestones Assessment and Placement Program) and powered by ChatGPT as the backbone for dialogue generation. Specifically, we designed intervention paradigms and prompts based on the clinical intervention method VB-MAPP and utilized ChatGPT's generative capabilities to facilitate social dialogue interventions. Experimental results demonstrate that our proposed system achieves competitive intervention effects to those of professional interventionists, making it a promising tool for long-term interventions in real healthcare scenario in the future.
\end{abstract}


\section{Introduction}
Autism Spectrum Disorder (ASD) is a neurodevelopmental disorder characterized by social communication deficits and repetitive, restricted behaviors or interests\cite{hirota2023autism}. According to a report, there is a global increase in the prevalence of ASD, with an estimation that 1 in 59 children are diagnosed with the disorder\cite{joudar2023artificial} worldwide. Intervention, as a method of treatment for children with ASD, is widely used in clinical scenarios. One of the popular intervention theories for ASD is Skinner's Verbal Behavior theory\cite{skinner1957verbal}, which categorizes verbal behavior into several verbal operants, such as mand, tact, echoic, intraverbal and so on. Referring to this theory, VB-MAPP (Verbal Behavior Milestones Assessment and Placement Program) is derived as an assessment tool widely used with individuals diagnosed with ASD or other language deficits\cite{sundberg2008vb}. Research indicates that 76\% of ASD-relevant professionals choose VB-MAPP as their assessment and intervention reference tool, highlighting its widespread use\cite{PADILLA2020101676}. However, due to the imbalance of medical resources and a shortage of professional healthcare providers, many children are unable to receive timely diagnosis and intervention, leading to lifelong struggles with ASD that continue to impact them into adulthood\cite{atherton2022autism}. Thus, the exploration of utilizing computer technology to alleviate the scarcity of medical resources for ASD intervention has remained a prominent focus in research. 

\begin{table*}
\label{tab:ref}
\caption{Comparison of the related work}
\begin{tabular}{lllllll}
\hline
Method & Group & Clinical Method & Personalized Customization & Clinical Validation & Intervention Type & Intervention Topic \\
\hline
Pedregal et al.\cite{pedregal2021autism} & Child & \XSolidBrush & \XSolidBrush & \Checkmark & Music & Emotion Recognition \\
Gagan et al.\cite{gagan2022designing}      &Adult       &\XSolidBrush &\Checkmark & \XSolidBrush  & Virtual Agent  & Understand Social Behavior              \\
Li et al.\cite{li2023faceme}       &Child       &     \XSolidBrush &\XSolidBrush &  \Checkmark   & AR+Virtual Agent             & Emotional Support                       \\
Wang et al.\cite{wang2020social}     &Child       &     \XSolidBrush &\XSolidBrush &  \Checkmark   & AR             & Communication Skills                    \\
Chung et al.\cite{chung2021robot}     &Child       &     \XSolidBrush &\XSolidBrush &  \Checkmark   & Robot          & Communication Skills                    \\
Holeva et al.\cite{holeva2024effectiveness}  &Child &     \XSolidBrush &\XSolidBrush &  \Checkmark   & Robot          & Psychosocial               \\
Perez et al.\cite{perez2024analysis}        &Child &     \XSolidBrush &\XSolidBrush &  \Checkmark   & Robot          & Emotional Support                       \\
Li et al.\cite{li2024exploring}          &Adult &     \XSolidBrush &\XSolidBrush &  \XSolidBrush & LLM+AR            & Communication Skills                    \\
Jang et al.\cite{jang2024s}       &  Adult        &     \XSolidBrush &\XSolidBrush &  \XSolidBrush & LLM & Emotional Support                       \\
Mishra et al.\cite{mishra2024towards}  &Child       &     \XSolidBrush &\Checkmark   &  \XSolidBrush & LLM+Robot            & Communication Skills                    \\
Tang et al.\cite{tang2024emoeden}    &Child       &     \XSolidBrush &\Checkmark   &  \Checkmark   & LLM            & Sentiment Control                       \\
\hline
\textbf{Ours}                      &Child       &     \Checkmark   &\Checkmark   &   \Checkmark  & LLM            & Communication Skills                    \\
\hline
\end{tabular}
\end{table*}

There is a plethora of research on computer-assisted interventions for ASD, including methods such as music\cite{pedregal2021autism}, virtual agents\cite{gagan2022designing}, augmented reality (AR)\cite{li2023faceme}, and robot-assisted therapy\cite{jain2020modeling}. Researchers design specific paradigms based on the intervention content and implement them using the entities mentioned above to facilitate the intervention. These methods provide children with multimodal stimuli that can be beneficial for their rehabilitation. However, the fixed paradigms can lead long-term interventions to become monotonous, resulting in a decrease in effectiveness. The impressive generation and conversational abilities of large language models (LLMs) can effectively address the limitations of customization in current methods. Therefore, recent studies have used LLMs as a backbone for designing intervention systems\cite{mishra2024towards,tang2024emoeden}. However, current methods have not taken into account the integration of real-world clinical intervention approaches such as VB-MAPP or ABLLS (Assessment of Basic Language and Learning Skills)\cite{usry2018using} to enhance the effectiveness of paradigms, making them more applicable to real intervention scenarios. To address the aforementioned problem, we propose an ASD dialogue intervention system called ASD-Chat. It uses VB-MAPP as the basis for paradigm and prompt design and then utilizes ChatGPT as the backbone for text generation. The main contributions of this paper are as follows:

\begin{itemize}
    \item Based on \textbf{VB-MAPP}, we have designed five paradigms related to ASD dialogue intervention, along with corresponding prompts to guide personalized conversation content.
    \item Based on the designed paradigms and ChatGPT, we propose an ASD dialogue intervention system called \textbf{ASD-Chat}. Experimental results indicate that ASD-Chat is competitive compared to professional interventionists through multi-modal data analysis.
\end{itemize}

\section{Related Work}
\subsection{Traditional Method for ASD intervention}
In recent years, researchers have used various computer technologies to aid in the intervention and treatment of ASD. Traditional methods typically employ stimuli from different modalities and design fixed paradigms targeting social interaction, emotions, and other aspects, to achieve independent or assisted interventions. Children with ASD often experience challenges in emotional control and recognition to some extent. One approach is to improve their ability to recognize emotional signals through music intervention, with the intervention effect evaluated using standardized scales\cite{pedregal2021autism}. Another approach involves using NAO robots to assist interventionists in emotional interventions\cite{holeva2024effectiveness}. Objective data such as eye contact, voice volume, and speaking time are collected to measure the effectiveness of the assisted intervention. The most typical symptom of individuals with ASD is social communication deficits. Gagan et al.\cite{gagan2022designing} addressed this by designing a virtual conversational agent named Amy to assist children in developing social communication skills. Amy utilizes voice and text interfaces, along with customized social story topics, to help children understand when to employ appropriate social behaviors. The effectiveness of robot\cite{perez2024analysis} or AR\cite{wang2020social} assisted interventions with interventionists has shown potential in this regard. Researchers aspire to further develop systems that can independently carry out interventions for ASD, thus conserving healthcare resources. So et al.\cite{so2023comparing} conducted a study on the application of robot drama in improving social communication skills among 38 Mandarin-speaking children, aged 6-9, with ASD. They found that robot intervention may be more effective than human intervention in promoting joined attention among autistic children with high support needs. FaceMe\cite{li2023faceme} uses AR and a virtual agent to teach basic facial expressions to children with ASD in social scenarios, with the aim of improving their emotional and communication skills. Clinical experiments have indicated that this system effectively stimulates positive social behaviors in children with ASD.

\subsection{ASD intervention via LLMs}
Traditional methods suffer from the limitation of fixed paradigms, which can lead to decreased intervention effectiveness due to monotony in long-term intervention processes. However, LLMs have garnered attention for their powerful generation and conversational capabilities and also show potential to the ASD intervention scenarios\cite{cho2023evaluating,jang2024s}. Li et al.\cite{li2024exploring} combined AR with LLMs and utilized virtual avatars to engage in voice conversations with ASD adults, aiming to improve their communication skills. However, they did not actually apply the system in clinical experiments, which means the effectiveness of the system remains to be validated. Jang et al.\cite{jang2024s} explored the opportunities and risks of LLM in adult patients with ASD. They had 11 individuals with ASD talk about their work-related social difficulties to (1) a GPT-4-based\cite{achiam2023gpt} chatbot and (2) a human assistant. The results showed that the participants strongly preferred interacting with the LLM rather than the human assistant. This further demonstrates the potential of LLMs in helping people with ASD in their daily lives\cite{choi2024unlock}. Mishra et al.\cite{mishra2024towards} integrated GPT-2 into NAO robots, enabling them to engage in verbal communication with children with ASD. The roles played by the robots, social scenarios, questions, and options were all customized through LLM-generated content, thereby avoiding the limitations imposed by fixed paradigms. However, similar to previous cases, the lack of clinical experiments leaves the effectiveness of the approach to be further examined and verified. EmoEden\cite{tang2024emoeden} is an emotion learning system designed for children with high-functioning autism (HFA). It integrates LLMs and text-to-image models, allowing virtual avatars on the interface to engage in conversations with the children. Additionally, based on information provided by children and parents, EmoEden can generate dialogue scenarios related to the child's interests, favorite foods, and more, in order to capture the child's attention and enhance its emotional tasks. The experimental results demonstrated significant improvements in emotional cognition and expression among the 6 participates with HFA. The positive effects persisted for five days after the user study.

The detailed comparison of related work is shown in Table \ref{tab:ref}. \textit{Clinical Method} indicates whether the system constructed in this work is designed based on the current clinical ASD intervention method, \textit{Personalized Customization} indicates whether personalized design has been added, and \textit{Clinical Validation} indicates whether the system has been clinically validated.

\begin{figure}
    \centering
    \includegraphics[width=1\linewidth]{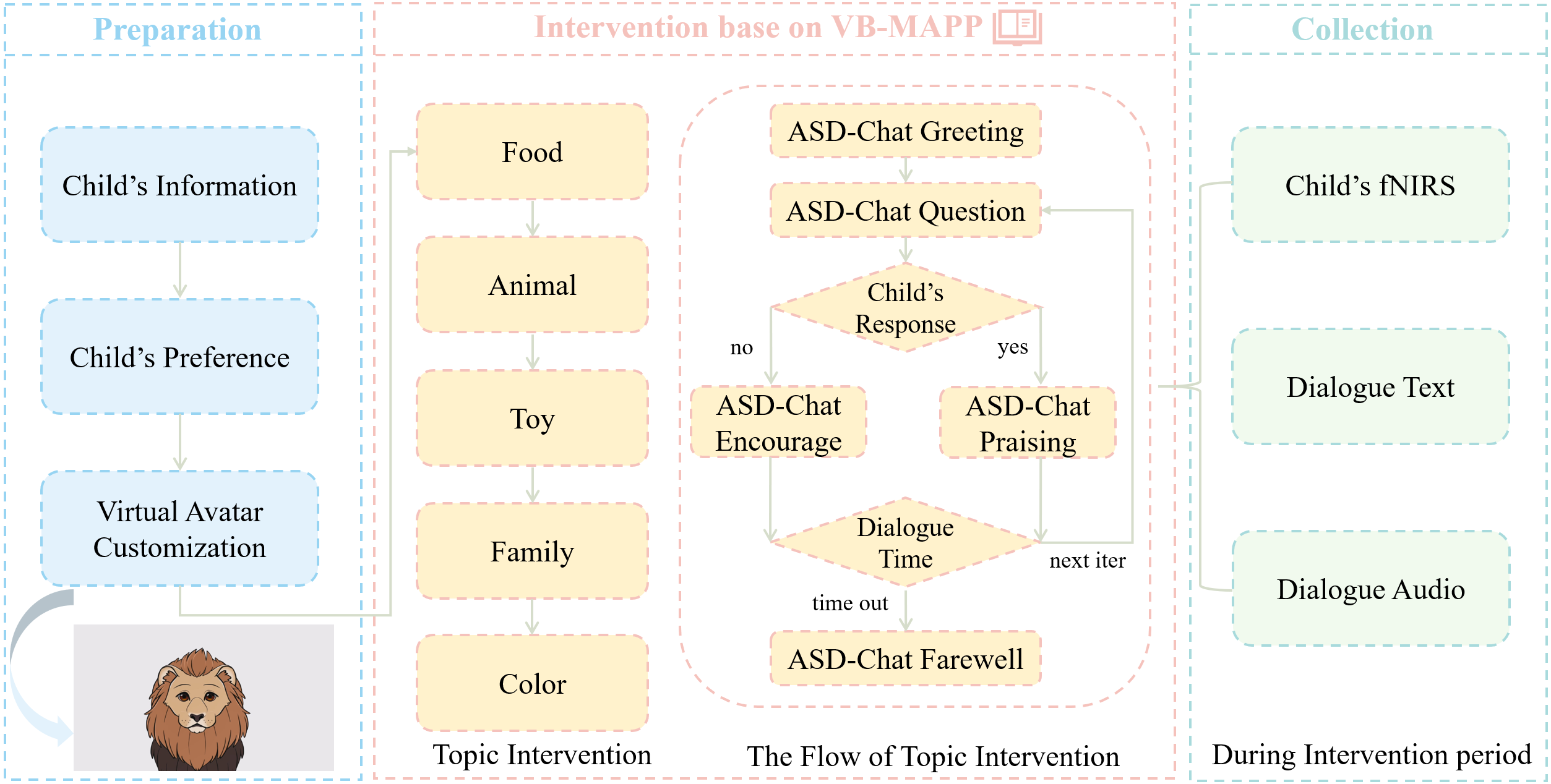}
    \caption{The Workflow of ASD-Chat. The complete flow of the conversation is divided into three stages: Preparation, Intervention, and Collection. In the Preparation stage, the child's basic information and preferences are inputted, and after selecting a virtual avatar, as a kind male lion is chosen and shown here, the Intervention stage begins. There will be five topic dialogue scenarios based on the VB-MAPP in the Intervention stage. ASD-Chat greets the child and initiates the formal dialogue intervention. Based on the child's responses, ASD-Chat provides encouragement or praise and then leading to a new round of dialogue. When a time-out occurs, ASD-Chat bids farewell to the child, marking the end of a dialogue topic. For Collection, data is collected synchronously during the conversation and saved sequentially when conversations finish.}
    \label{fig:asd-chat}
\end{figure}

\section{Method}

\subsection{Paradigm Design Based on VB-MAPP}

VB-MAPP is an assessment tool widely used with individuals diagnosed with ASD or other language deficits\cite{PADILLA2020101676}. The assessment is used to evaluate performance on Skinner’s verbal operants\cite{skinner1957verbal} across a number of tasks. For example, there are five tasks under the intraverbal part of the second developmental level (equivalent to the degree of normal 18- to 30-month-old children), one of them is "\textit{Answer 25 different questions about what}", with detailed scoring criteria: "\textit{If the child can answer 25 different 'what' questions without echoic prompts or the presentation of related objects, a score of 1 is given. If the child can answer 12 such questions, a score of ½ is given}". VB-MAPP provides some instructions and objections on intervention for individuals based on the comprehensive situation of various scores. For children who scored low in the example task ahead, the focus of intervention suggestions is "expanding and generalizing this ability". Depending on these instructions and objectives, interventionists can make moderate intervention approaches. For individuals with ASD, 
turn-taking conversation is a typical and effective form of intervention recommended by VB-MAPP. Based on the discrete trial training guidelines provided by VB-MAPP and the recommendations of experienced interventionists, we designed a turn-taking conversation paradigm centered around predefined topics and restricted formats. There are five predefined topics involved: food, animal, toy, family, and color. These topics are arranged in order of increasing cognitive difficulty. In the conversation procedure, based on the sections in VB-MAPP related to intraverbal and the opinions of interventionists who are familiar with VB-MAPP, the question form is constrained to the field of what, who, and where, excluding why, how-to and when. The what, who, and where questions link to specific visual images, making them easy for children with social communication deficits to answer. But, answering why, how, and when questions pose a real challenge for individuals with ASD. To address these questions, they need better logical thinking, abstract generalization, and time perception ability, which is difficult for them. Moreover, this can lead to frustration and decreased engagement, reducing the effectiveness of the intervention among individuals who already struggle with communication.

\begin{figure}
    \label{tab:system-prompt}
    \centering
    \includegraphics[width=1\linewidth]{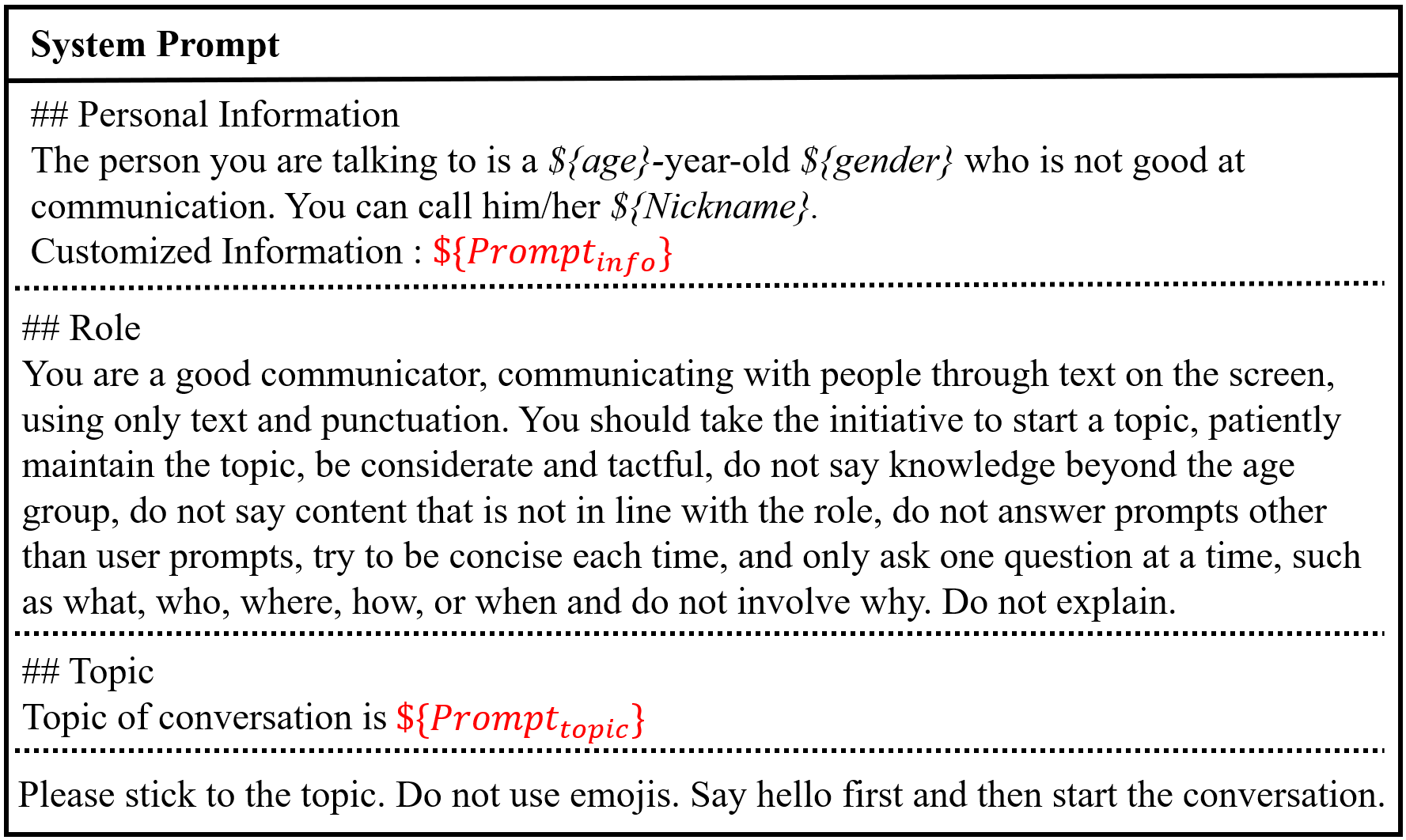}
    \caption{Template of system prompt used by ASD-Chat. This prompt is inputted into ASD-Chat at the beginning of the conversation, allowing it to understand the basic information about the child, the current topic context, and its basic conversational restrictions. \textcolor{red}{\(Prompt_{info}\)} indicates the child's preference such as food and color. \textcolor{red}{\(Prompt_{topic}\)} indicates the detail description of current topic.}
    \label{fig:sys-prompt}
\end{figure}

\subsection{ASD-Chat System}
To imitate the turn-taking dialogue intervention conducted by human interventionists, we design and construct a multi-modal computer program system called ASD-Chat. In abstract terms, the system has three modules: sensation, performance, and control of the virtual agent. In the sensation module, the subsystem gathers voice and other signals about the subject. The necessary part of the sensation module is the speech voice collection, where the text spoken by the subject is produced. The performance module creates a profile that can be perceived by the subject. With the selected modalities from voice, visual avatar, feedback animation, and so on, the subject will be treated by a cordial virtual agent, and feel pleasant accepting the intervention. As for the control module, a set of rules will control the workflow and data flow of the system so that the system acts like an interventionist who is conducting a turn-taking conversation intervention. The gathered information flows through some data handling programs under the control of rules, produce the instructions for performance module. 

The intervention process of the system is depicted in Fig. \ref{fig:asd-chat} and consists of three stages: Preparation, Intervention, and Collection. In the Preparation stage, we input the child's information and preferences through the interface. This information is then injected into the system prompt to personalize the dialogue. Once the virtual avatar is selected, we proceed to the Intervention stage. In the Intervention stage, the procedure of each conversation around the topic can be described as follows: The virtual agent makes a greeting and introduces the topic first, the subject has enough time to answer or keep silent. While the time reaches the limit, the AI-driven virtual agent responds to the dialogue and continues the conversation based on the topic and some constraints which will be explained later in this article. The turn-taking conversation will keep going until the time for this topic is exhausted, and the virtual agent summarizes the conversation and farewell to the subject. The subject has the final opportunity to say goodbye, then the conversation is finished. The Collection stage spans the entire Intervention stage and we gather functional near-infrared spectroscopy (fNIRS), dialogue text, and audio generated during the intervention. Once the intervention is finished, these data are sequentially stored for further analysis.

LLM is a significant part of this system. We implement OpenAI's ChatGPT to receive the text of what the subject said and other control instructions and generate the text of what the virtual agent should say next. With others covered by API, we can focus on the parameters provided by OpenAI, especially the text inputs to LLM which are also referred to as "prompts". We program the initial system prompt carefully so that the LLM can serve as an ideal character who can chat with subjects in a friendly tone. An example is given as Fig. \ref{fig:sys-prompt}, with the text translated from Chinese. The initial prompt can be divided into three parts. The first part is the basic and related information about the subject, which can be compiled as "Personal Information". \textit{\(Prompt_{info}\)} contains the information of the subject related to the topic, such as preferences, interests, and recent experiences. ChatGPT can learn from it and generate personalized and customized responses. The second part contains the role and how to play the role in detail. The last part is the topic of this conversation, \textit{\(Prompt_{topic}\)} contains the keyword and several entry points. For the food topic, the entry points are pets, knowledge about animals, and what if you become an animal. Besides the initial system prompt, there are more control instructions in the following prompts. Also, they are all translated from Chinese. If the subject remains silent during the should-speak time duration, a description of this situation will be added to the prompt; if the subject says something but the speech recognition (SR) server provides no text, the system gives a default description like "\emph{Unrecognized speech, duration approximately \$\{speech\_time\} second(s).}" to ChatGPT. If the time of conversation reaches its limit, there will be a special system prompt that says "\emph{The communication time has ended, respond to my words, summarize our communication, and say goodbye to me.}". When none of the above situations occur, the system prompt is "\emph{Continue communicating around the topic and the content mentioned before. Ask only one question at a time, such as what, who, and where, without involving why, how-to, or when. Don't explain. If the speech is unclear, you can ask for a repeat.}", which aims to emphasize the paradigm mainly. All these prompts are generated by the closed-loop system itself. With more feedback devices involved, more feedback prompts can be added depending on the situation.

\section{Experiment}
\subsection{Experiment Paradigm}
To validate the effectiveness of the designed paradigms and system, we have devised the following experimental paradigms based on the hypothesis that if the system achieves significant intervention outcomes, certain data metrics from children's conversations with the system should closely resemble the results obtained from professional ASD interventionists. Therefore, we conducted conversations with 12 children who were diagnosed with ASD and were determined by their interventionists to have certain language abilities, and our proposed system, ASD-Chat. The conversations were centered on five topics designed based on VB-MAPP, with each topic lasting about 3 minutes. Before conversations, we collect preferences from parents of children, such as their favorite foods, to personalize the system prompt. In addition, we have professional interventionists engage in conversations with the children on the same five topics, keeping the total duration within 15 minutes. By comparing the results with those obtained from the ASD-Chat system, we can investigate the effectiveness of ASD-Chat.

For the ASD-Chat system, several implementation details are shown here. The skeleton of system is programmed by Python.  The text-to-speech function supported by Microsoft's Azure Cognitive Services makes virtual agent speaking possible. SR function is supported by iFLYTEK open platform's real-time ASR (automatic speech recognition) service which can recognize the voice of Chinese children with low language ability. For LLMs, we choose ChatGPT (gpt-3.5-turbo-0125) as the LLM server, for it's a basic and typical choice. The virtual agent was given an appearance of a male lion as shown in Fig. \ref{fig:asd-chat} with a kind adult male voice. The animated virtual agent can react to special situations if triggered, such as greeting while the conversation starts, nodding while subjects speak, and speaking while playing the audio synchronously.


\subsection{Data Collection \& Pre-Processing}
12 ASD children(11 males, 1 female) aged between 4 and 11 years old comprised the experiment sample whose base information is shown in Table \uppercase\expandafter{\romannumeral2}. ASD diagnosis was confirmed by ADOS-2 (Autism Diagnostic Observation Schedule, Second edition)\cite{lord2012autism}, and participants were required to have basic communication abilities judged by interventionists. The Human Subjects Committee of the Children's Hospital of Zhejiang University School of Medicine approved this study.

Three types of data during the intervention were collected during the experiment, namely text, audio, and fNIRS signals. The text and audio from the dialogue interventions with ASD-Chat are automatically saved by the system and the audio with the interventionists is recorded using a recording pen (XXJ1, HP Inc., California, USA) saved as .wav and transcribed into text, with speaker identification annotations using Paraformer\cite{gao2022paraformer} and Cam++\cite{wang2023cam++}. To ensure the accuracy of the transcribed text, we performed manual calibration by making corrections based on the model's recognition. Speech segments are divided from audio as preprocessing, with outliers discarded.

fNIRS is a noninvasive neuroimaging technique that monitors changes in the concentrations of oxyhemoglobin (HbO) and deoxyhemoglobin (HbR), which are associated with activities of the cerebral cortex\cite{villringer1997non}. Here we only analyzed HbO changes, which generally change more significantly during brain activity and are therefore easier to detect and analyze\cite{hoshi2007functional}. fNIRS signals were acquired using an 8-channel fNIRS system (NirSmart, Huichuang Corp., Danyang, China) with a frequency resolution of 30 Hz. The anatomical locations of optodes were placed near the ventromedial prefrontal cortex (vmPFC), which is in the inhibition of emotional responses, and the process of decision-making and self-control\cite{hiser2018multifaceted}. Data preprocessing was completed using the Homer2 package in MATLAB (R2023a). The raw intensity data were first converted to optical density (OD) changes. Then motion artifacts were detected and spline interpolated. A FIR filter (0.01–0.2 Hz) was used to extract the low-frequency fluctuations, and the OD data were converted to hemoglobin concentration changes using the modified Beer-Lambert law\cite{kocsis2006modified}. The first 1 second of oxygenated hemoglobin (HbO) data before the conversation was used for steady-state control and normalized by z-score methods to eliminate the effect of data units and facilitate comparison between different conditions. Finally, each channel's HbO amplitude under the five dialogue topics of each subject in the two task scenarios was calculated to compare the similarity of the two scenarios.

\renewcommand{\arraystretch}{1.3}
\label{tab:base-info}
\begin{table}
\centering
\begin{threeparttable}
\caption{Participant demographic information}
\begin{tabularx}{80mm}{l@{\hspace{15mm}}>{\raggedright\arraybackslash}X>{\raggedright\arraybackslash}l}
\hline
Descriptive variables & Mean & SD \\
\hline
Age & 6.26 & 1.92 \\
Sex & 11M, 1F & - \\
ADOS & 9.08 & 3.26 \\
-Social Communication & 6.83 & 2.67 \\
-Restricted/Repetitive Behaviors & 2.25 & 1.54 \\
\hline
\end{tabularx}
\begin{tablenotes}
\footnotesize
\item M = Male, F = Female.
\end{tablenotes}
\end{threeparttable}
\end{table}

\section{Analysis}
To validate the effectiveness of the ASD-Chat system, this study conducts verification from two aspects: social communication behavior and physiological signals.

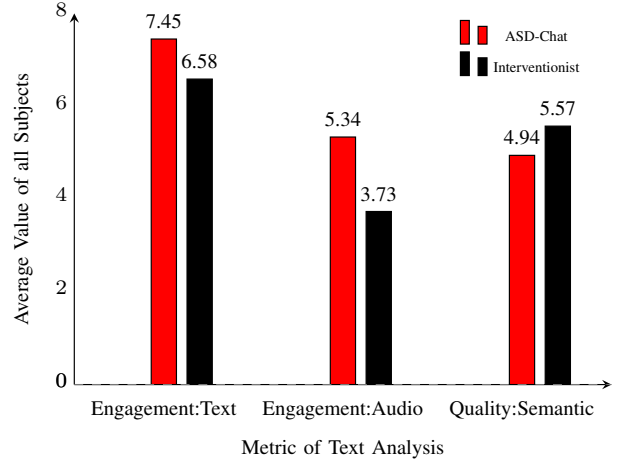
\begin{figure}[t]
    \centering
    \begin{adjustbox}{width=0.9\columnwidth, trim=0.1cm 0 0 0, clip} 
        \begin{tikzpicture}
            \begin{axis}[
                ybar,
                bar width=0.3cm, 
                xtick={1,2,3, 4, 5}, 
                xticklabels={Engagement:Text, , Engagement:Audio, , Quality:Semantic}, 
                ymin=0, 
                ymax=8, 
                xmin=0, 
                xmax=6, 
                width=0.9\columnwidth, 
                height=6cm, 
                xlabel={\scriptsize Metric of Text Analysis}, 
                ylabel={\scriptsize Average Value of all Subjects}, 
                axis on top,
                xlabel style={font=\scriptsize}, 
                every axis y label/.style={at={(ticklabel cs:0.5)}, rotate=90, anchor=south, font=\scriptsize, yshift=0cm, inner sep=0pt}, 
                yticklabel style={font=\scriptsize, inner sep=0pt, yshift=0.5mm}, 
                tick label style={font=\scriptsize}, 
                xtick style={draw=none},
                ytick style={draw=none},
                axis x line=bottom, 
                axis y line=left, 
                ymajorgrids=false, 
                extra y ticks=0, 
                extra y tick style={grid=major, grid style={line width=0.25pt, color=gray, dashed}}, 
                xlabel near ticks,
                ylabel near ticks,
                every node near coord/.append style={font=\tiny}, 
                legend style={at={(0.7,1)}, anchor=north west, legend columns=1, font=\tiny,draw=none} 
            ]

            \addplot[
                ybar,
                bar shift=0cm, 
                fill=red
            ] coordinates {
                (1,7.447) (3,5.335) (5,4.940)
            }node[pos=0,above, font=\scriptsize] {7.45}
            node[pos=0.5,above, font=\scriptsize] {5.34}
            node[pos=1,above, font=\scriptsize] {4.94};

            \addplot[
                ybar,
                bar shift=0cm, 
                fill=black
            ] coordinates {
                (1.4,6.584) (3.4,3.730) (5.4,5.570)
            }node[pos=0,above, font=\scriptsize] {6.58}
            node[pos=0.5,above, font=\scriptsize] {3.73}
            node[pos=1,above, font=\scriptsize] {5.57};
        
            \addlegendentry{ASD-Chat}
            \addlegendimage{area legend, fill=red}
            \addlegendentry{Interventionist}
            \addlegendimage{area legend, fill=black}
            \end{axis}
        \end{tikzpicture}
    \end{adjustbox}
    \vspace{-0.3cm}
    \caption{The effectiveness of the ASD-Chat system in social intervention scenarios is evaluated from two aspects: \textit{Engagement} and \textit{Quality}. \textit{Engagement} is measured by the average length of pure text and speaking duration per dialogue round, while \textit{Quality} is assessed based on the semantic coherence in each dialogue round.}
    \label{fig:text-bar}
\end{figure}

\begin{table*}[]
\centering
\label{tab:text-audio}
\caption{Comparison between ASD-Chat and interventionist of Speech}
\begin{tabular}{cclrrrrrrrrrrrrr}
\hline
field                          & metrics              & \multicolumn{1}{c}{type} & \multicolumn{1}{c}{1} & \multicolumn{1}{c}{2} & \multicolumn{1}{c}{3} & \multicolumn{1}{c}{4} & \multicolumn{1}{c}{5} & \multicolumn{1}{c}{6} & \multicolumn{1}{c}{7} & \multicolumn{1}{c}{8} & \multicolumn{1}{c}{9} & \multicolumn{1}{c}{10} & \multicolumn{1}{c}{11} & \multicolumn{1}{c}{12} & \multicolumn{1}{c}{avg} \\ \hline
\multirow{4}{*}{\begin{tabular}[c]{@{}c@{}}speech\\ sound\end{tabular}}  & \multirow{2}{*}{F0}  & ASD-Chat                 & 687.3                 & 656.8                 & 789.4                 & 545.8                 & 666.4                 & 581.3                 & 614.6                 & 656.3                 & 807.9                 & 616.2                  & 783.1                  & 449.9                  & 655                     \\ \cline{3-16} 
                               &                      & Interventionist                   & 656.9                 & 458.6                 & 750.9                 & 591.1                 & 821.6                 & 465.7                 & 759.9                 & 620.8                 & 752.4                 & 753.9                  & 681.5                  & 485.1                  & 650                     \\ \cline{2-16} 
                               & \multirow{2}{*}{ZCR} & ASD-Chat                 & 0.042                 & 0.037                 & 0.038                 & 0.039                 & 0.049                 & 0.044                 & 0.043                 & 0.042                 & 0.039                 & 0.028                  & 0.029                  & 0.013                  & 0.037                   \\ \cline{3-16} 
                               &                      & Interventionist                   & 0.040                 & 0.015                 & 0.031                 & 0.020                 & 0.027                 & 0.044                 & 0.026                 & 0.024                 & 0.026                 & 0.024                  & 0.021                  & 0.016                  & 0.026                   \\ \hline
\multirow{10}{*}{\begin{tabular}[c]{@{}c@{}}voiced\\ sound\end{tabular}} & \multirow{2}{*}{F0}  & ASD-Chat                 & 724.3                 & 633.0                 & 977.6                 & 536.8                 & 670.5                 & 593.9                 & 626.9                 & 669.6                 & 873.5                 & 738.3                  & 911.8                  & 452.4                  & 701                     \\ \cline{3-16} 
                               &                      & Interventionist                   & 683.6                 & 484.1                 & 741.3                 & 544.1                 & 754.2                 & 468.6                 & 695.6                 & 569.1                 & 746.0                 & 715.8                  & 639.3                  & 477.3                  & 627                     \\ \cline{2-16} 
                               & \multirow{2}{*}{ZCR} & ASD-Chat                 & 0.043                 & 0.035                 & 0.042                 & 0.040                 & 0.049                 & 0.046                 & 0.045                 & 0.042                 & 0.042                 & 0.030                  & 0.032                  & 0.013                  & 0.038                   \\ \cline{3-16} 
                               &                      & Interventionist                   & 0.038                 & 0.020                 & 0.031                 & 0.021                 & 0.028                 & 0.056                 & 0.028                 & 0.026                 & 0.028                 & 0.026                  & 0.022                  & 0.017                  & 0.029                   \\ \cline{2-16} 
                               & \multirow{2}{*}{F1}  & ASD-Chat                 & 485                   & 474                   & 502                   & 453                   & 435                   & 413                   & 438                   & 465                   & 464                   & 476                    & 485                    & 732                    & 485                     \\ \cline{3-16} 
                               &                      & Interventionist                   & 655                   & 493                   & 672                   & 516                   & 608                   & 456                   & 579                   & 536                   & 572                   & 582                    & 563                    & 468                    & 558                     \\ \cline{2-16} 
                               & \multirow{2}{*}{F2}  & ASD-Chat                 & 1163                  & 1170                  & 1213                  & 1218                  & 1216                  & 1262                  & 1241                  & 1226                  & 1194                  & 1254                   & 1265                   & 1586                   & 1251                    \\ \cline{3-16} 
                               &                      & Interventionist                   & 1290                  & 1123                  & 1339                  & 1145                  & 1255                  & 1185                  & 1226                  & 1220                  & 1187                  & 1133                   & 1124                   & 1141                   & 1197                    \\ \cline{2-16} 
                               & \multirow{2}{*}{F3}  & ASD-Chat                 & 2106                  & 2134                  & 2119                  & 2145                  & 2125                  & 2134                  & 2138                  & 2134                  & 2094                  & 2108                   & 2130                   & 2329                   & 2141                    \\ \cline{3-16} 
                               &                      & Interventionist                   & 2116                  & 2086                  & 2124                  & 2091                  & 2088                  & 2184                  & 2095                  & 2109                  & 2081                  & 2085                   & 2073                   & 2093                   & 2102                    \\ \hline
\end{tabular}
\end{table*}


\subsection{Social Communication Behavior : Text \& Speech}
We primarily evaluate the behavioral performance of children with ASD during intervention from two aspects: text and speech. By comparing their performance with that of professional interventionists, we aim to demonstrate that the proposed ASD-Chat system is competitive. The results of the text analysis are shown in Fig. \ref{fig:text-bar}, we evaluate the effectiveness from two aspects: \textit{Engagement} and \textit{Quality}.

\textit{Engagement} reflects the level of child participation during the intervention process. The higher the engagement, the higher the child's level of enthusiasm. It also indirectly reflects that ASD-Chat is capable of capturing the child's attention and achieving the desired intervention effect. The level of engagement can be measured by the length of the text and the duration of the audio. We collected data on the average number of words (excluding punctuation) and the average duration of each child's sentences within the five topics. From Fig. \ref{fig:text-bar}, it can be observed that when facing ASD-Chat, the children spoke more words and had longer durations compared to when facing professional interventionists. On average, the number of words spoken when interacting with ASD-Chat was 13.11\% higher, and the duration was 43.03\% longer than when interacting with professional interventionists. One possible explanation is that children may find it easier to engage in communication with a virtual cartoon character compared to interacting with adults. This is because they are accustomed to engaging in similar dialogues while playing with toys in their daily lives. Communicating with a virtual character may put them in a more relaxed state, thereby increasing their enthusiasm and active participation during the conversation.

\textit{Quality} reflects the quality of the child's responses during the intervention dialogue. This metric aims to investigate whether children engage in active participation by thoughtfully considering and providing correct responses to the ASD-Chat system or professional interventionists, rather than simply speaking without logical reasoning. We used the method of semantic similarity to measure the quality of the child's responses. Based on the following assumption: relevant question-answer pairs should have a close proximity in semantic space, we located the child's response and extracted the preceding sentence from ASD-Chat or the interventionist as a text pair for calculation. We inputted these sentence pairs into the BGE-M3\cite{chen2024bge} text retrieval model to vectorize the sentences and calculate their cosine similarity to indicated the semantic relevance. In Fig. \ref{fig:text-bar}, the \textit{Quality} displays the average semantic similarity scores for each participant across the five topics. A higher score indicates a better quality of responses. We have to acknowledge that in terms of the quality of responses, the interventionists performed slightly better, with ASD-Chat scoring 11.31\% lower. The interventionists' extensive clinical intervention experience allows them to easily guide children to provide correct answers through language and actions. However, the current ASD-Chat system lacks this ability and is still in need of improvement in this aspect. 

At the audio level, the analysis focuses on the subject's speech sound. Speech segments are divided from audio as preprocessing, with outliers discarded. Then the voiced sound segments are classified by referring to the frequencies' patterns of formants\cite{schafer1970system}. The fundamental frequency (F0) and zero cross rate (ZCR) can, to some extent, represent the speech characteristics of specific individuals, and the level of arousal of specific individuals\cite{banse1996acoustic}. The average F0 and ZCR of speech sound and voiced sound are listed in Table \uppercase\expandafter{\romannumeral3} with the first three formants' frequencies of voiced sound. F1 is related to the opening degree of vowels, F2 is related to the anterior and posterior positions of the tongue, and F3 is related to the position of the tongue tip and lip shape. In statistical terms, due to the different frequencies of various voiced sounds in the comparative experiments, there are differences in the means of F0, ZCR, F1, F2, and F3. Similar variations occur in the speech sound as well. It can be seen that on these metrics, the performance of the same subject is close to that with interventionists and with the ASD-Chat system. Specifically, the F0 of speech sound in the experiment with the ASD-Chat system was 2\% higher than with interventionists. The ZCR of speech sound was 48\%, and the ZCR of voiced sound was 40\%. There are two possible influencing factors: the image of the virtual human may cause tension in the subjects; and different audio acquisition devices were used in the comparative experiment. In the voiced sound metrics, F0 was 12\% higher, F1 was 11\% lower, F2 was 5\% higher, and F3 was 2\% higher, In summary, they are very close.

However, we can still see the potential of the ASD-Chat system. It has the ability to induce children to produce contextually appropriate responses within social topics using simple visual and language stimuli, without the need for involvement from other professionals. In the future, it can be deployed as an independent intervention tool in low-resource areas, enabling automated ASD interventions. This would allow more ASD children to enhance their social communication skills through social dialogue interventions at a low cost.

\begin{figure}[t]
    \centering
    \includegraphics[width=0.8\columnwidth]{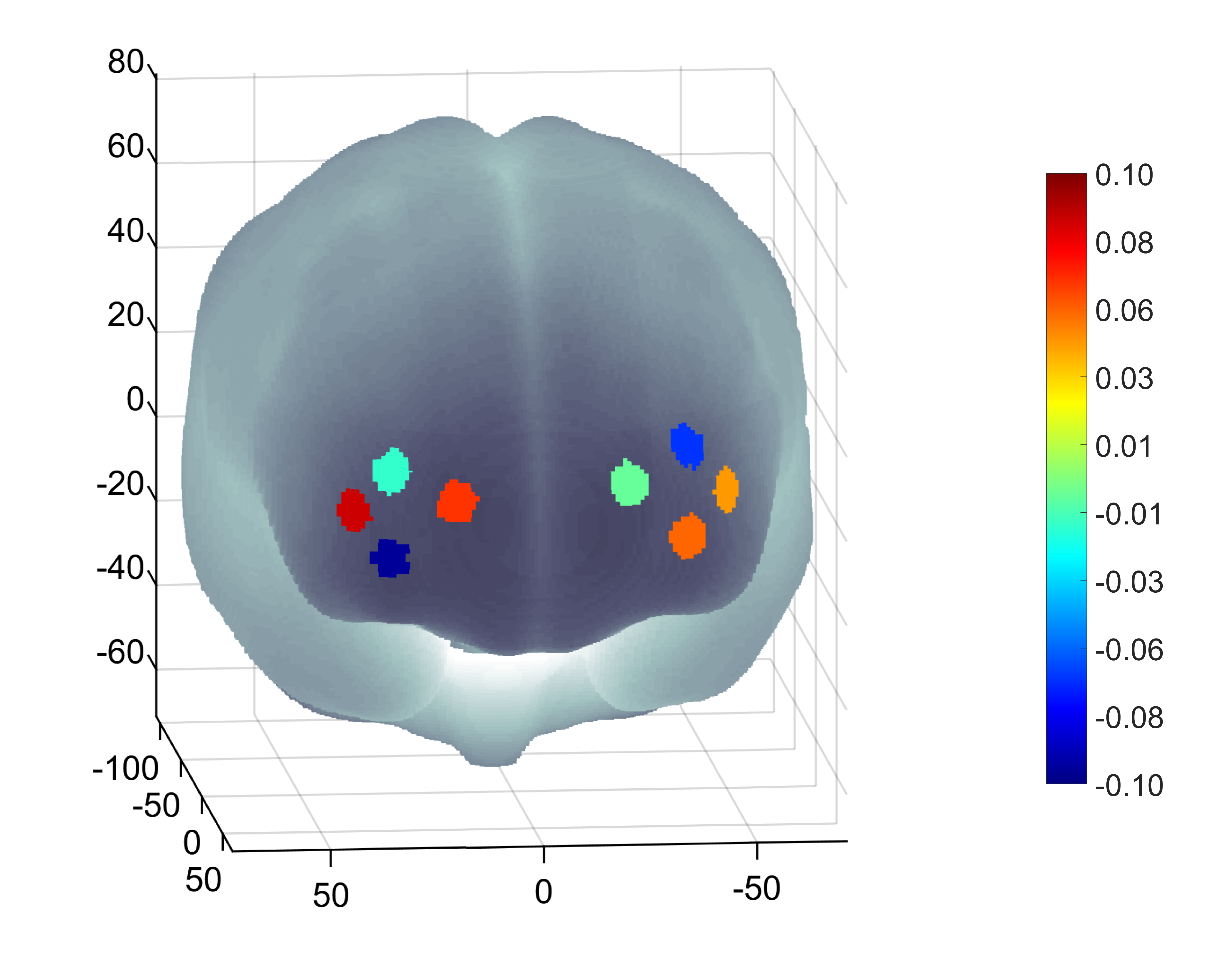}
    \caption{The interpolation of the subject-averaged HbO amplitude for each channel across all subjects. Both scenarios have strong brain activations in different channels, but the magnitude of this difference is around 0.1, meaning that the two scenarios have similar brain activation effects.}
    \label{fig:topography}
\end{figure}

\subsection{Physiological signals: fNIRS}
We used HbO amplitude evoked by the same topic to illustrate the similarities between the ASD-Chat system and the interventionist scenario. The blood oxygen amplitude was especially linked to social and communication features, representing the core symptoms of ASD\cite{mazziotti2022amplitude}. As shown in Fig. \ref{fig:topography},  we calculated the difference in HbO amplitude for each channel of all autistic children in the five topic conversations of two scenarios. Compared with the interventionist scenario, half of the channels in the ASD-Chat scenario had greater HbO amplitude changes, while the brain activation intensity of the four channels was weaker. Notably, even though the interventionist scenario conversation caused relatively stronger fNIRS signal changes on some channels, the range of the difference between the two was only around 0.1. The difference between the two was almost negligible, which means that there was almost no difference in the HbO amplitude changes caused by the two scenarios. It proved that the ASD-Chat intervention system proposed in this paper has similar effects to the traditional intervention method.

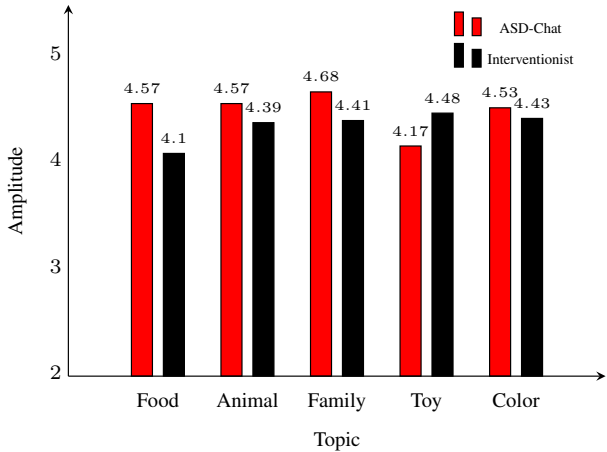
\begin{figure}[t]
    \centering
    \begin{adjustbox}{width=0.9\columnwidth, trim=0.1cm 0 0 0, clip} 
        \begin{tikzpicture}
            \begin{axis}[
                ybar,
                bar width=0.25cm, 
                xtick={1,2,3,4,5}, 
                xticklabels={Food, Animal, Family, Toy, Color}, 
                ymin=2, 
                ymax=5.5, 
                xmin=0, 
                xmax=6, 
                width=0.9\columnwidth, 
                height=6cm, 
                xlabel={\scriptsize Topic}, 
                ylabel={\scriptsize Amplitude}, 
                axis on top,
                xlabel style={font=\scriptsize}, 
                every axis y label/.style={at={(ticklabel cs:0.5)}, rotate=90, anchor=south, font=\scriptsize, yshift=0cm, inner sep=0pt}, 
                yticklabel style={font=\scriptsize, inner sep=0pt, yshift=0.5mm}, 
                tick label style={font=\scriptsize}, 
                xtick style={draw=none},
                ytick style={draw=none},
                axis x line=bottom, 
                axis y line=left, 
                ymajorgrids=false, 
                extra y ticks=0, 
                extra y tick style={grid=major, grid style={line width=0.25pt, color=gray, dashed}}, 
                xlabel near ticks,
                ylabel near ticks,
                nodes near coords,
                every node near coord/.append style={font=\tiny}, 
                legend style={at={(0.7,1)}, anchor=north west, legend columns=1, font=\tiny, draw=none} 
            ]
            
            \addplot[
                ybar,
                bar shift=-0.19cm, 
                fill=red
            ] coordinates {
                (1,4.57) (2,4.57) (3,4.68) (4,4.17) (5,4.53)};
        
            \addplot[
                ybar,
                bar shift=0.19cm, 
                fill=black
            ] coordinates {
                (1,4.10) (2,4.39) (3,4.41) (4,4.48) (5,4.43)};

            \addlegendentry{ASD-Chat}
            \addlegendimage{area legend, fill=red}
            \addlegendentry{Interventionist}
            \addlegendimage{area legend, fill=black}
            \end{axis}
        \end{tikzpicture}
    \end{adjustbox}
    \vspace{-0.3cm}
    \caption{The average amplitude of all channels in each topic conversation for both scenarios of Subject 9. The ASD-Chat scenario has stronger brain activations for most topics (food, animal, family, and color). Although the toy topic of the interventionist scenario evokes larger amplitude changes, the interpolation between the two is small, only 0.31, which means that the activation strength is similar.}
    \label{fig:barChart}
\end{figure}

\begin{figure}[t]
    \centering
    \includegraphics[width=\columnwidth]{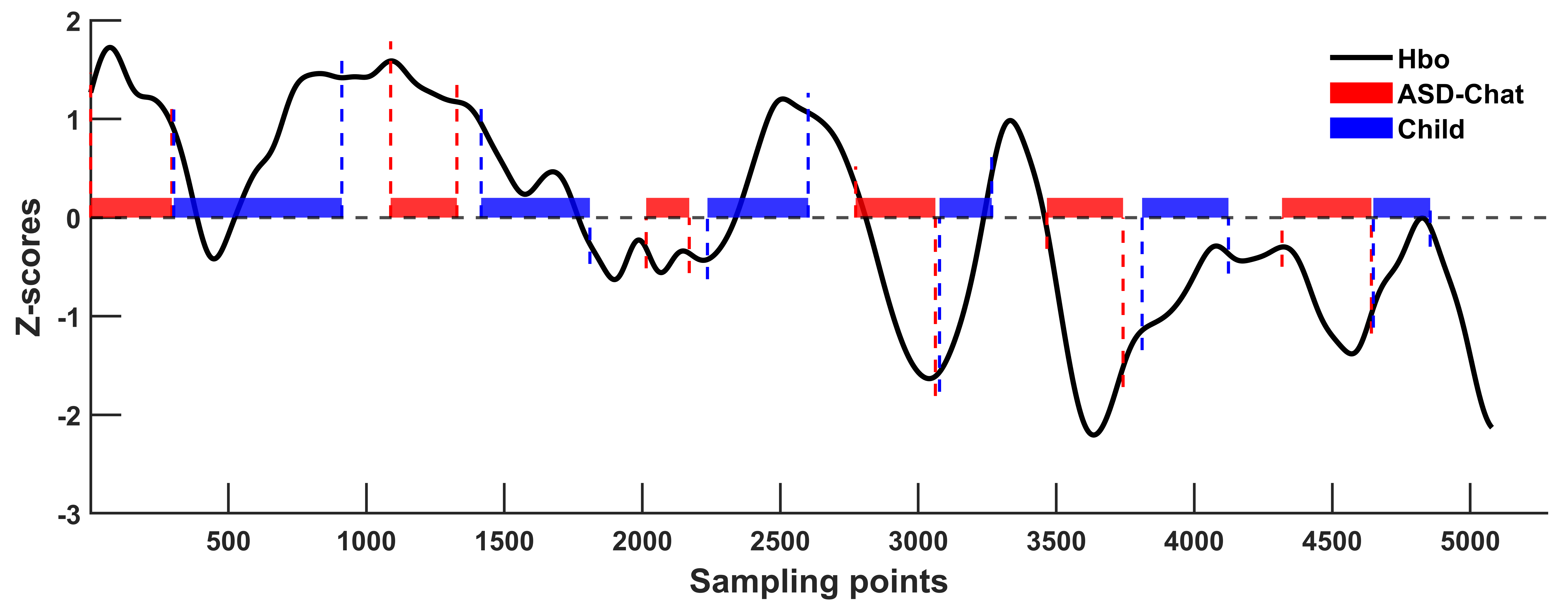}
    \caption{The HbO waveform changes of channel 4 under the color topic in the ASD-Chat scenario of Subject 9. After the ASD-Chat system conversation, HbO showed a typical hemodynamic response, and the duration of each round of conversation of the subjects was longer than that of the ASD-Chat system.}
    \label{fig:brain topography}
\end{figure}

\subsection{Case Study}

We select subject 9 as a case study. Fig. \ref{fig:barChart} shows the Channel-average HbO amplitude in five topics of two scenarios. Red represents the ASD-Chat scenario and black represents the interventionist scenario. Four topics(food, animal, family, color) have greater HbO amplitude changes and stronger brain activation in the ASD-Chat scenario except for the toy topic. While the scenario of the interventionist has a larger amplitude in the toy topic, the amplitude difference between the two scenarios is only 0.31 (ASD-Chat: 4.17, interventionist: 4.48). In the food topic conversation, the fNIRS signal activation of the ASD-Chat scenario is significantly better than that of the interventionist scenario, and the amplitude difference between the two scenarios reaches 0.47 (ASD-Chat: 4.57, interventionist: 4.1). In the color topic conversation, there is almost no difference between the two scenarios. In general, the ASD-Chat scenario may cause stronger brain activation in ASD children and is expected to effectively improve the social language conversation ability of ASD children at the individual level.

We also visualized the changes in the HbO waveform of channel 4 of the subject in the color theme dialogue as shown in Fig. \ref{fig:brain topography}. The red interval represents the virtual agent speaking, and the blue interval represents the subject speaking. After the virtual agent spoke, the subject's HbO showed a typical hemodynamic response, and the brain was significantly activated. In terms of the topic duration, the entire topic conversation lasted about 166.67 seconds, the subject's speech time was about 69.20 seconds, and the ASD-Chat system's speech time was about 52.63 seconds. It indicated that subjects can actively engage in the intervention process, and the ASD-Chat system can effectively stimulate relevant brain region activities.


\section{Conclusion}
We designed a paradigm and prompts based on VB-MAPP and built the ASD-Chat system for dialogue intervention in ASD children, with ChatGPT as the backbone. Through a comparative experiment with professional interventionists, we have demonstrated the effectiveness of the system through data analysis from three modalities: text, speech audio, and fNIRS.

The aforementioned analysis demonstrates the potential of the ASD-Chat system, as it can achieve intervention effects similar to those of professional interventionists. Future work will primarily focus on two aspects. Firstly, the current experiments mainly concentrate on short-term interventions. Subsequent evaluations will assess the changes before and after interventions from a long-term perspective to further validate the effectiveness of the system. This will enable its practical implementation in real healthcare settings and alleviate resource scarcity issues. Secondly, we will enrich the stimulus paradigms of the current system by incorporating more modalities such as visual question-answering, making the intervention methods more diverse and personalized.

\bibliographystyle{unsrt}
\bibliography{reference}



\end{document}